\documentclass[apj]{emulateapj}
\pdfoutput=1

\usepackage{graphicx}
\usepackage{amsmath}
\usepackage{amssymb}
\usepackage[colorlinks=false,pdfborder={0 0 0}]{hyperref}
\usepackage{natbib}
\usepackage[]{threeparttable}

\begin{document}

\title{Ice Grain Collisions in Comparison: CO$_2$, H$_2$O and their Mixtures}

\author{Grzegorz Musiolik, Jens Teiser, Tim Jankowski, Gerhard Wurm}
\email{}

\address{Fakult{\"a}t f{\"u}r Physik, Universit{\"a}t Duisburg-Essen, Lotharstr. 1, 47048 Duisburg, Germany}

\begin{abstract}
Collisions of ice particles play an important role in the formation of planetesimals and comets. In recent work we showed, that CO$_2$ ice behaves like silicates in collisions. The resulting assumption was that it should therefore stick less efficiently than H$_2$O ice. Within this paper a quantification of the latter is presented. We used the same experimental setup to study collisions of pure CO$_2$ ice, pure water ice and 50\% mixtures by mass between CO$_2$ and water at 80K, 1 mbar and an average particle size of $\sim 90 \mu$m.
The results show a strong increase of the threshold velocity between sticking and bouncing with increasing water content. This supports the idea that water ice is favorable for early growth phases of planets in a zone within the H$_2$O and the CO$_2$ iceline.

\end{abstract}

\section{Introduction}

Ices are important constituents in the collisional formation of comets and planetesimals in protoplanetary disks.
As various species of ice appear in different distances to the central star due to their individual sublimation pressures, there are various icelines within the disk. Planet formation in general but especially collision outcomes are tied to these icelines and the physics of the prevailing kind of ice.  \citep{Alidib2014,Aumatell2011,Aumatell2014,Deckers2016,Blum2015,Musiolik2016}.

In earlier experiments we showed that it is necessary to distinguish the different ices while studying their growth potential \citep{Musiolik2016}.  We suggested that collisional growth should be most efficient between the water and the CO$_2$ iceline or between 2.0 to 9.3 AU according to the minimum mass solar nebula model. Beyond 9.3 AU non-polar CO$_2$ dominates. Collision experiments showed that CO$_2$ collisions are comparable to silicate collisions. An increased sticking of water ice was proposed but not verified in experiments. 

H$_2$O is often considered as a driving mechanism for planetesimal growth due to a high sticking efficiency compared to other materials \citep{Gundlach2011}. While this is plausible as water does have a high electrical dipole moment and as individual experiments with water ice point to this \Citep{Gundlach2015}, a direct experimental comparison with other ices under similar conditions is missing. Therefore, we compare collisions of pure CO$_2$, pure H$_2$O and 1:1 mixtures by mass here.

\section{The Experiment}

\subsection{Experimental setup}

\begin{figure}[t!]
	\centering
	\includegraphics[width=\columnwidth]{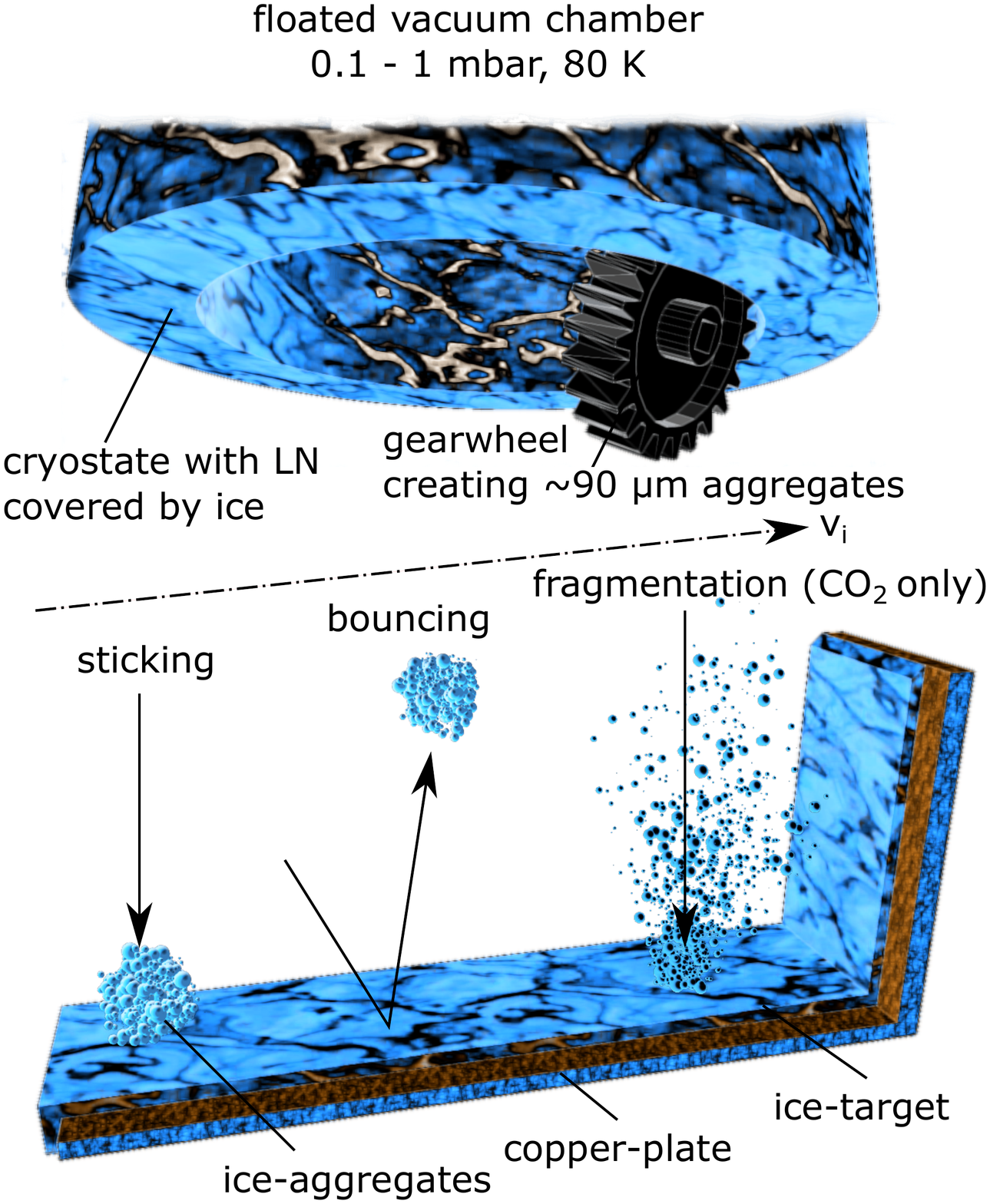}
	\caption{\label{kryostat}
		Experimental setup: A vacuum chamber is flooded with a gas (mixture) which is deposited on a cryostate cooled with liquid nitrogen. Then the chamber is evacuated to an ambient pressure of 0.1-1 mbar and particles are beveled off by using a rotating gearwheel. The particles collide with an ice layer on a copper plate connected to the cryostate. Depending on the impact velocity, sticking or bouncing can be observed. In earlier work we also observed fragmentation for CO$_2$ \citep{Musiolik2016}.
	}
\end{figure}

The experimental setup is equal to the former experiment from \citet{Musiolik2016}. Details can be found there. In short, the experiment consists of a vacuum chamber flooded with different gas samples. In the former experiment we used CO$_2$-gas. In this work we additionally take a mixture between CO$_2$ and H$_2$O gases and H$_2$O vapor. A scheme of the experimental setup is shown in fig. \ref{kryostat}.

The chamber is cooled by liquid nitrogen to a temperature of 80 K in the collision section. Within several minutes a 2 mm thick ice layer is deposited on the walls of the chamber. 
The chamber is sealed and evacuated to an ambient pressure of $\sim$1 mbar. By means of a gearwheel driven by an electrical motor ice grains are beveled off from the cryostate. This produces $\sim$90 $\mu$m ice particles. The chemical composition of the produced aggregates depends on the gas (mixture) which was used to float the chamber. The particles collide with an ice-layer of the same composition which is deposited on a copper plate connected to the cryostate. Collision velocities reach up to 1 m/s in this work. 
Collisions are imaged with a camera at 1250 frames per second with a spatial resolution of 10 $\mu$m.

\subsection{Sample preperation and size measurement}

The colliding ice particles are generated by scraping them off from the cooled cryostate. The structure and the chemical composition of these aggregates depends on the initial atmosphere within the vacuum chamber. 

For pure CO$_2$ aggregates the chamber is just continuously flooded by a stream of CO$_2$ gas. The CO$_2$ sublimation point is at about 195 K \citep{Mazzoldi2008}. Therefore, we argue that the molecules only freeze out once they hit a cold wall but do not already form ice grains within the gas phase. The scraping then removes material from a solid ice surface. According to our earlier analysis of fragmenting collisions the CO$_2$ particles are small aggregates. 

For the H$_2$O-CO$_2$ aggregates we use the same procedure with a mixture of CO$_2$ gas and H$_2$O vapor. The mass ratio between CO$_2$ and H$_2$O is 1:1. This value was determined after the experiment by weighing the mass loss of the ice mixture outside of the chamber, which is caused by the rapid evaporation of CO$_2$. Unlike the CO$_2$ vapour the water supply already consists of $\sim\mu$m water droplets besides vapour. The structure of the ice surface will be slightly different then. We currently cannot analyze this in detail but would argue on the state of the ice surface as follows:

Some of the water droplets will diffuse towards the walls and stick there as water ice grains. However, gas molecules will reach the surface by diffusion preferentially (due to their smaller mass compared to the droplets) and stick there as in the pure CO$_2$ case. This leads to an intimate mixture of CO$_2$ and water ice molecules with a certain amount of pure water ice grains embedded which can be regarded as homogeneous material.
Our collision results are in favor of this view. As seen below, we see a clear threshold velocity between bouncing and sticking for particles from the mixed sample.  This threshold is distinguished from the pure CO$_2$ threshold by an order of magnitude. Would the particles which are beveled off be inhomogeneous and consist of large fractions of pure water or pure CO$_2$ ice or otherwise locally distinguished compositions at their surface, then an individual collision should depend on the material during contact which should e.g. be 
water or CO$_2$ in a given collision. In any case we should not see a clear threshold in collision experiments then.  The same argument holds if there is a significant amount of ice 
grains embedded but if 
scraping preferentially breaks the mixed matrix in between which is supposedly less stable. Also then only one sort of mixed surface
would interact with the same material. In this case the mixing ratio might not be the mixing ratio of the matrix material measured (1:1) though, but might be shifted towards the CO$_2$ fraction. As the CO$_2$ fraction is high
in this case the mix will glue together any pure water ice spheres efficiently, though. They might be porous, but the aggregates cannot restructure at the given impact energies and
can be considered as individual grains for low velocities.

For the H$_2$O sample we flooded the chamber only with water vapour and droplets. The same arguments hold that there should be a mixture of water ice droplets embedded in a water ice matrix. It is likely though that the water ice matrix is very thin and weak.
In this case it might not be appropriate to treat the aggregate particles as single grains as already small 
impact velocities might be sufficient to restructure the aggregate. Some elasticity in larger aggregates 
is clearly visible in the high speed observations for the case of pure water ice. This changes the outcome
of collisions (e.g. \citep{Dominik1997}). It will depend on the specific configuration now and the amount of
energy that is dissipated by restructuring if an aggregate sticks or bounces. We should not expect
a clear threshold velocity then.

\begin{figure}
	\centering
	\includegraphics[width=\columnwidth]{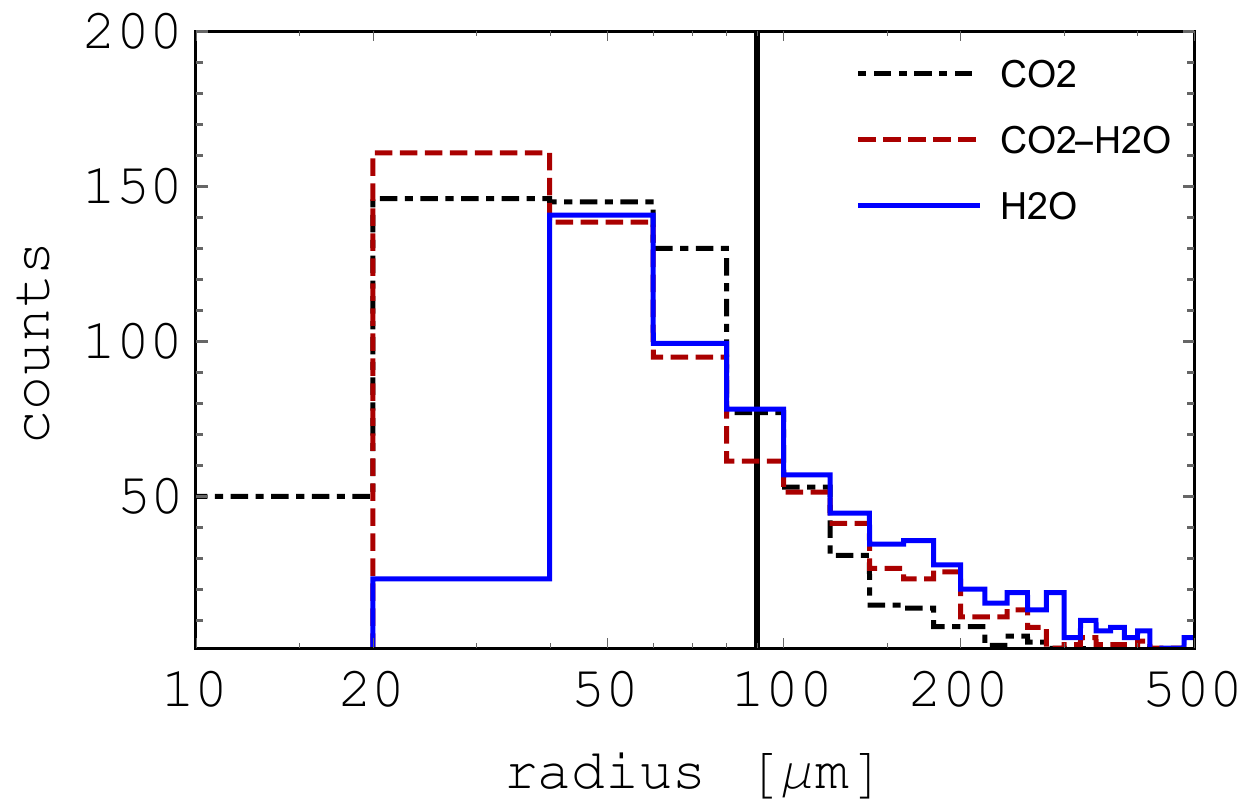}
	\caption{\label{size}
		Size distrubution for the different species of ice particles considered for this study. The distribution for the CO$_2$ aggregates is taken from \citet{Musiolik2016}. The black line marks the average size of 90 $\mu$m. A bin size of 20 $\mu$m is chosen with respect to the spatial resolution of the camera.}
\end{figure}

We consider the pure CO$_2$ particles as mixed particles as solid spheres even if they would be aggregates. All particles are initially modeled as spheres of a size with the same cross sections as the observed grains. We determined the cross section by optical imaging. This way we get size distrubutions of the particles which can be found in fig. \ref{size}.
These distributions are comparable; they all peak around $\sim$50 $\mu$m and have a medium size of $\sim$90 $\mu$m. The bin size of 20 $\mu$m is chosen due to the spatial resolution of the optical system of 10 $\rm \mu m$.  There is an indication in the size distribution of a cut-off at 20 $\rm \mu m$. In any case, \citet{Krijt2014} showed that there is a smallest fragment in a dissipative process so it is not likely for all particles to be covered by a layer of small unseen grains. Smaller particles may be present but should not affect the observed collision outcomes.

The analized grain sizes are a narrow but representative illustration of dust found in protoplanetary disks. Here, the sizes reach from sub-$\mu m$, which is a typical size for dust in the ISM \citep{Williams2011,Draine2003}, to the cm-regime at which the bouncing barrier sets in \citep{Zsom2010}. The size distribution typically follows a power law \citep{Draine2006}.

Since the distributions of all analyzed paticles (fig. \ref{size}) are comparable we can directly compare the collisional behavior for one mean size of the different species of ice grains depending on the collision velocity
keeping in mind though that aggregation might affect the result.

\section{Results and Discussion}

Collision outcomes at low speed can be categorized in two types: 1) sticking between the grain and the target at low velocities; 2) bouncing from the target at higher impact velocities \citep{Zsom2010,Heisselmann2010,Ormel2007,Blum2000,Weidenschilling1993}. The threshold between sticking and bouncing is not only important for understanding whether collisional growth of aggregates in protoplanetary disks is possible but also allows to determine fundamental parameters like the surface energy of the particles \citep{Dominik1997}. In general, there are more effects for higher impact velocities like the fragmentation of or mass transfer to the aggregates \citep{Blum2008,Deckers2016} which we do not consider in this work.

\begin{figure*}
	\centering
	\includegraphics[width=0.8\textwidth]{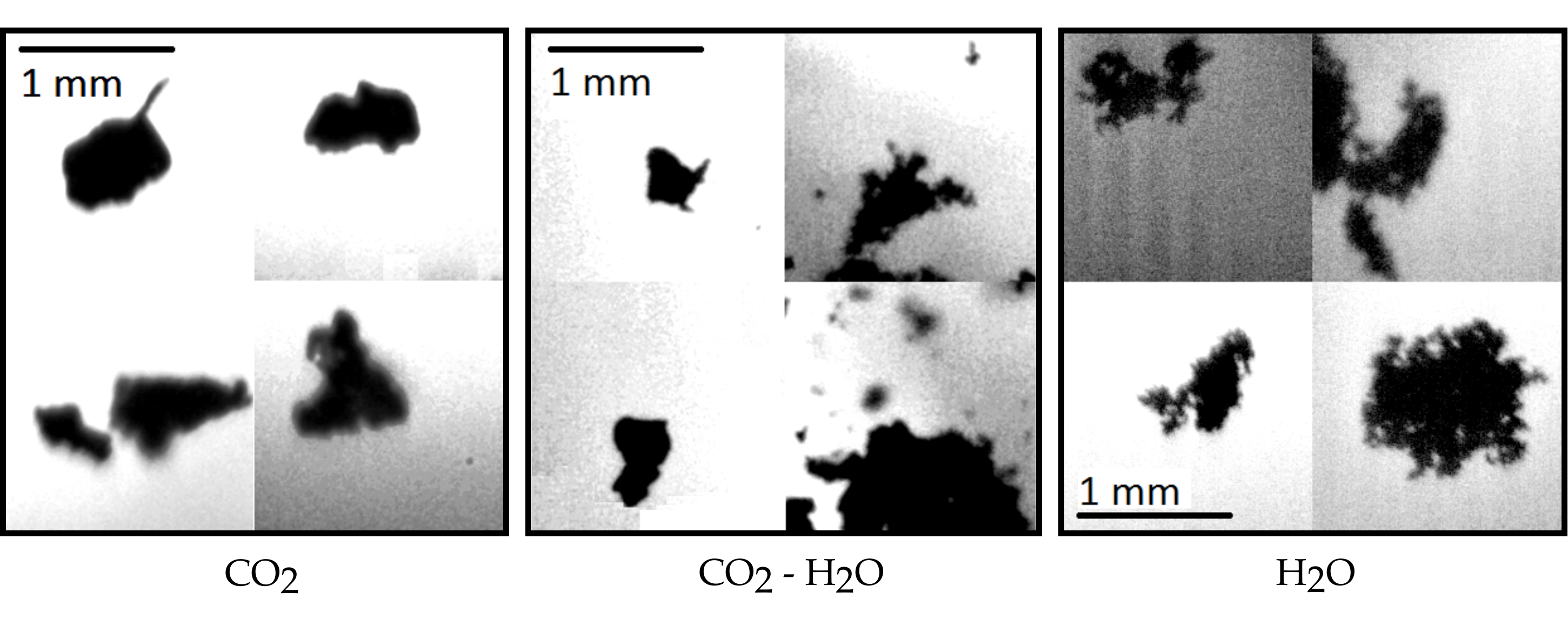}
	\caption{\label{surface}
		The structure of the largest aggregates from the dataset for the various compositions. For each ice species four aggregates are shown to illustrate the internal structures. The  morphology of the water ice shows a high porosity in contrast to more compact morphologies otherwise.}
\end{figure*}

In fig. \ref{surface} we show  samples of the largest aggregates for each dataset of different ices (not analyzed collision-wise). For the largest aggregates, the surface structure and porosity can be determined easiest. Observation of these particles suggest a more irregular surface structure and higher porosity for water ice. A quantitative analysis of the surface and the porosity is not possible so far. As described above, due to the variation in the microstructure between the CO$_2$-aggregates and the CO$_2$-H$_2$O-/ H$_2$O-aggregates a comparision of their collisional behavior is given only roughly.

The porous structure of the water ice aggregates suggests that restructuring might be important as indicated above. The mixed ice samples qualitatively look compact enough that they should only show hit-and-stick or just bouncing slightly above the threshold for sticking. 

\begin{figure}
	\centering
	\includegraphics[width=\columnwidth]{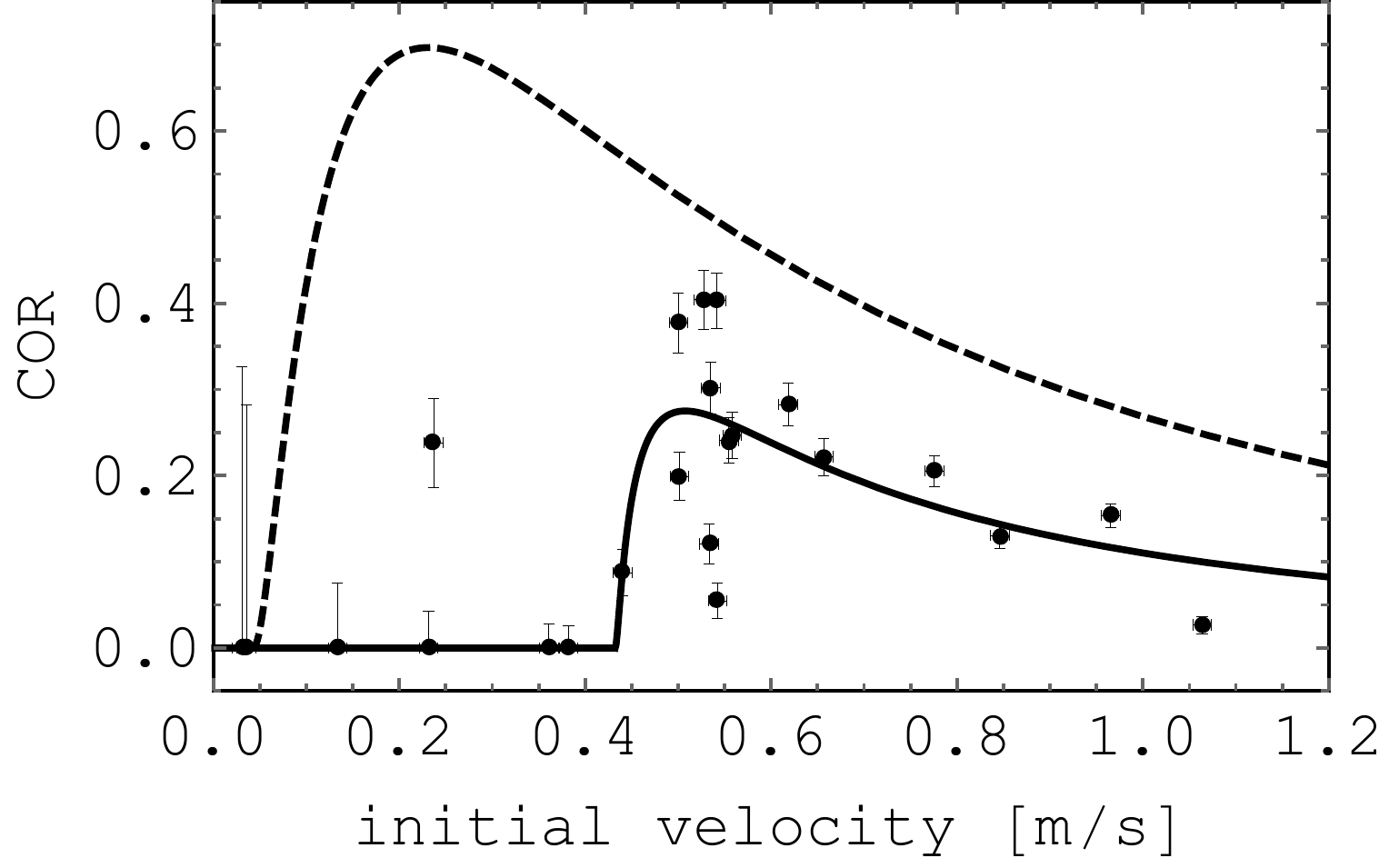}
	\caption{\label{COR_MIX}
		The coefficient of restitution (COR) for the mixture particles with a H$_2$O-CO$_2$ mass-ratio of 1:1. The dashed line describes the COR for pure CO$_2$ particles from \citet{Musiolik2016}. The solid line represents the fit function from eq. \eqref{restmodel}. The uncertainties result from uncertainties in the impact velocity and the aggregate size.}
\end{figure}

More quantitatively, the collisional behavior can be described by the coefficient of restitution (COR) which we define as the ratio of the absolute value of the particle velocity after $v_\text{o}$ and before $v_\text{i}$ the collision
\begin{equation}
	\epsilon = \frac{v_\text{o}}{v_\text{i}}.
\end{equation} 
There are many theroretical models describing the COR for elastic, plastic and viscous spheres \citep{Andrews1930,Thornton1998,Tomas2006,Antonyuk2010}. Other approaches try to describe the COR by heuristic functions, like \citet{Higa1996,Higa1998} did for water ice spheres. 
For the COR resulting from the observed collisions we use our model deduced in the earlier work on pure CO$_2$ collisions \citep{Musiolik2016}.
\begin{equation}
\epsilon(v_\text{i}) = A\cdot e^{\left(a_1\left(\ln\left(\frac{v_\text{i}-v_\text{stick}}{v_\text{c}}\right)\right)^2 \right)}\Theta(v_\text{i}-v_\text{stick})
\label{restmodel}
\end{equation} 
with the critical velocity $v_\text{c}$, the sticking velocity $v_\text{stick}$ and material dependend parameters $a_\text{1}$ and $A$. Here, the sticking velocity $v_\text{stick}$ describes the transition between the sticking and the bouncing regime, where $\epsilon(v\leq v_\text{stick})=0$ and $\epsilon(v>v_\text{stick})>0$. The critical velocity $v_\text{c}$ describes the maximum for the COR, where the bouncing is most elastic. The COR from eq. \eqref{restmodel} models three different effects; sticking with $\epsilon(v\rightarrow0)\rightarrow 0$, elastic bouncing with $\max(\epsilon(v)) =\epsilon(v_\text{c})$ and plastic deformation with $\epsilon(v\rightarrow \infty)\rightarrow 0$. This function gives the most appropriate fit for the COR for CO$_2$ particles. It also fits the COR for the H$_2$O/CO$_2$ mixture well like shown in fig. \ref{COR_MIX}.
The fit parameters for both fits are summarized in tab. \ref{fitpar}.

\begin{table}[t]\centering
	\begin{threeparttable}
	\caption{Fit parameters for the COR.}
	\label{fitpar}
	
	\begin{tabular}{ccc}\hline\hline
		& CO2   & CO$_2$-H$_2$0 \\ \hline
		$A$                        & 0.67 $\pm$ 0.04  & 0.27 $\pm$ 0.05   \\
		$a_1$                      & -0.36 $\pm$ 0.04 & -0.22 $\pm$ 0.4   \\
		$v_\text{c}$ {[}m/s{]}     & 0.189 $\pm$ 0.025 & 0.075 $\pm$ 0.12   \\
		$v_\text{stick}$ {[}m/s{]} & 0.04 $\pm$ 0.02  & 0.43 $\pm$ 0.03  \\ \hline\hline
	\end{tabular}
	
	\begin{tablenotes}
		\small \centering
		\item The values for CO$_2$ particles are taken from \citet{Musiolik2016}.
	\end{tablenotes}
	\end{threeparttable}
\end{table}

The gravitational force is acting on the aggregates during collisons. This effect distorts the analysis of the data, because low COR collisions might be classified as sticking events. Nonetheless, in \citet{Musiolik2016} we show, that this effect becomes significant only for collisions with impact velocities below 0.05 m/s. This is also the reason for deducing the sticking velocity for CO$_2$ aggregates by a model fit in the previous work. 

Collisions of CO$_2$-H$_2$0 particles are more inelastic than collisions of CO$_2$, since the COR in fig. \ref{COR_MIX} has smaller values for the mixture. Moreover the sticking velocity is an order of magnitude larger. From the sticking velocity we can calculate the surface energy for CO$_2$-H$_2$0 particles from \citep{Dominik1997},
\begin{equation}
v_\text{stick} =  \frac{1.07}{\rho^{1/2}E_\text{py}^{1/3}} \cdot   \frac{1}{R^{5/6}} \cdot  \gamma^{5/6} ,
\label{stickvel}
\end{equation}
with the surface energy $\gamma$, the reduced radius $R$, the particle mass density $\rho$ and an elastic constant $E_\text{py}$, where $E_\text{py}=E_\text{y} / (2(1-\nu^2_\text{p}))$. In this definition $E_\text{y}$ is the Young's modulus and $\nu_\text{p}$ is the Poisson's ratio. 

Taking a mean Young's modulus between water ice and CO$_2$ ice of $E_\text{y}=1/2(13+9)$ GPa, a mean density of $\rho=1/2(1000+1560)$ Kg/m$^3$ and a mean poisson ratio of $\nu_\text{p}=1/2(0.3+0.28)$ \citep{Musiolik2016,Yamashita1997, Nimmo2004} we get a surface energy for particles with a mean radius of $r_p=90 \pm 20 \mu$m of 
\begin{equation}
	\gamma_\text{mix}=2.77^{+0.9}_{-0.8}  \text{J}/\text{m}^2.
	\label{senergy}
\end{equation}

We determined the surface energy for pure CO$_2$ ice in our earlier work to $\gamma_\text{CO2}=0.17  \text{J}/\text{m}^2$ \citep{Musiolik2016}.
Compared to this the surface energy of mixed particles this is an order of magnitude higher. 

\begin{figure}
	\centering
	\includegraphics[width=\columnwidth]{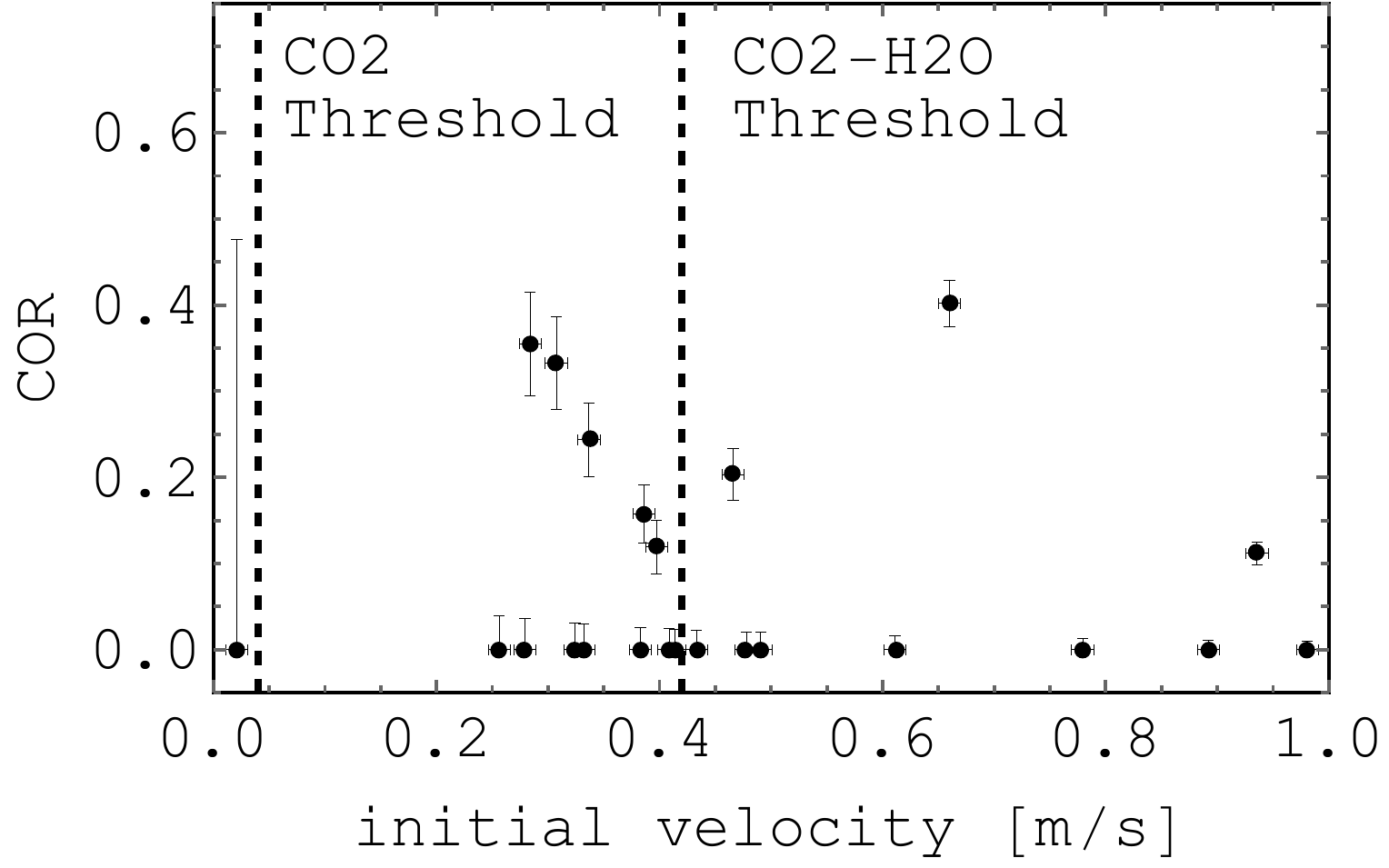}
	\caption{\label{COR_W}
		The coefficient of restitution (COR) for  water ice particles. The dashed lines describe the threshold for CO$_2$ and CO$_2$-H$_2$O data.}
\end{figure}

For the H$_2$O aggregates, we do not see a sharp transition from the sticking regime to the bouncing regime which might be due to the preparation and restructuring of the aggregates in collisions. Fig. \ref{COR_W} shows the measured data for the coefficient of restitution.
We can determine a sticking probability $P$ though, defined as the ratio between all sticking events in collisions $N_\text{s}$ and all analyzed collison events $N$ for  certain velocity range.
\begin{equation}
P=\frac{N_\text{s}}{N}.
\label{stickingprob}
\end{equation}

For impact velocities between 0-0.5 m/s the sticking probability is $P=0.65$ and for impact velocities between 0.5-1 m/s $P=0.67$. 

\begin{figure}
	\centering
	\includegraphics[width=\columnwidth]{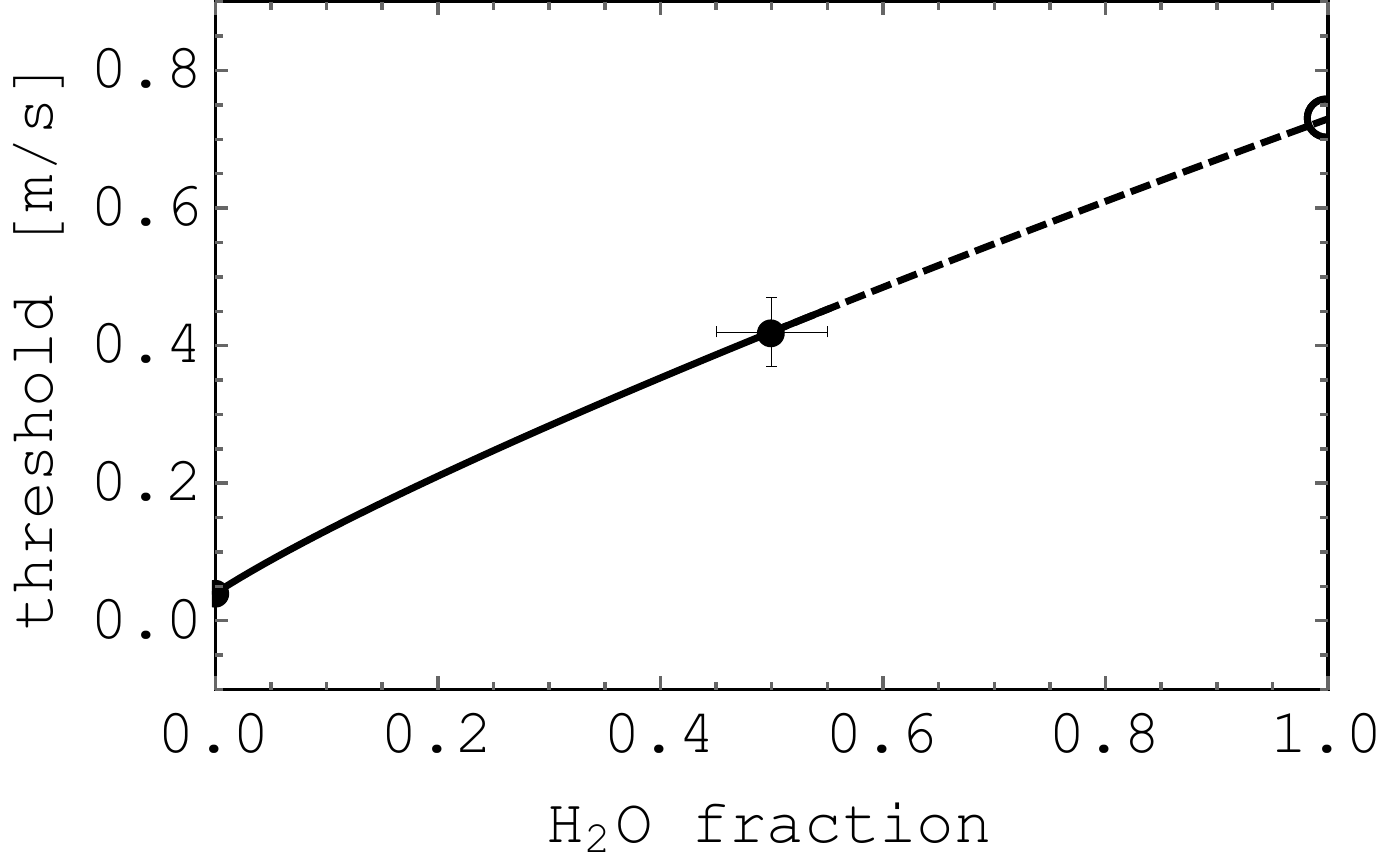}
	\caption{\label{Threshold}
		Threshold velocities between the sticking and the bouncing regime depending on the H$_2$O fraction of the ice. The black dots show measured values and the circle an estimation from eq. \eqref{vpw}. The black line shows the fit function which changes over to the extrapolation.}
\end{figure}

For comparability of pure water ice, one can use eq. \eqref{stickvel} to get a rough estimate of the dependency between the H$_2$O fraction $p_W$ in the particles and the sticking velocity $v_\text{stick}(p_W)$ for the range $p_W \in [0,1]$.
Assuming a simple linear approximation of $\gamma \propto p_W + \xi_1$ with the constants $\xi_1, \xi_2$ we obtain
\begin{equation}
v_\text{stick}(p_W) =\xi_2 \left(p_W+\xi_1\right)^{5/6},
\label{vpw}
\end{equation}
Using eq. \eqref{vpw} we can fit the sticking velocities determined for $p_W=0$ and $p_W=0.5$ and extrapolate this function for an H$_2$O fraction of 1. For the constants we get $\xi_1= 0.031$ and $\xi_2=0.711$ m/s. This procedure leads to a value of $v_\text{stick}(1)\approx 0.73$  m/s (fig. \ref{Threshold}). Within the range studied we do see sticking and bouncing of pure water ice at this velocity. The measured water data is therefore at least not in contradiction to this extrapolation. 

In their recent work \citet{Gundlach2015} give a sticking threshold for 1.5 $\mu$m sized water ice of 9.6 m/s. If we scale this value with the dependency from eq. \eqref{stickvel}, $v_\text{stick} \propto R^{-5/6}$, we get a sticking velocity of 0.31 m/s for 90 $\mu$m pure water ice grains. Within the range of unknowns (porosity, regular - irregular grain relation) our data are also consitent with this.

\section{Caveats}

Fig. \ref{Threshold} clearly shows that threshold velocities for sticking of H$_2$O/CO$_2$ mixtures vary between the less sticky end member CO$_2$ and the much stickier other end of pure H$_2$O ice. 
The analytical dependence between $\gamma$ and $p_W$ chosen -- or in general the sticking velocity dependence on $p_W$ -- could be more complex though. In general, the JKR-Theory can give us only an approximation here anyway, because it does not describe nonspherical aggregate shapes and complicated contacts including dipole moments or hydrogen bonds. 
A more detailed analysis based on thorough variation of the mass ratios ($p_W$) would be
needed to quantify this further. 
However, as outlined above the preparation does not warrant homogeneous mixtures yet but grains might e.g. be core-mantle particles with embedded water ice grains. In this case
the total mass ratios determined by sublimation of the CO$_2$ afterwards could be different from mass ratios of the surfaces which are actually in contact during a collision.
Hence, such a study would require a spatially resolved analysis of the composition of the 
grains on the nm-scale. This seems feasible with current microscopic techniques but due to the volatile nature of the ices this is not straightforward and is beyond the scope of this paper.

We used the sticking velocity measured and modeled as given and deduced the surface energy based on
eq. \ref{stickvel}. 
This might lead to a mismatch with given literature data. Values for the surface energy of pure H$_2$O ice are e.g. 0.37 J/m$^2$ from \citet{Hirashita2013} or 0.19 J/m$^2$ from \citet{Gundlach2011,Blum2015}. Our value of the mixed sample should be below the water value and
the determined value of 2.8 J/m$^2$ looks to be too large then (by an order of magnitude). Our deduced values for pure CO$_2$ is also overestimating existing values by a factor of a few \citep{Musiolik2016}.
Experimental work from \citet{Blum2000, Poppe2000} suggested that the model by \citet{Dominik1997} might give values for $v_\text{stick}$ which are an order of magnitude too low. If this is a theoretical (model) / experimental mismatch, our deduced values from eq. \eqref{stickvel} could be overestimated by an order of magnitude though they would be self-consistent among themselves. Another uncertainty in the calculation of the surface energy is the structure of the particles. For once they are irregular and not spheres and likely can be small aggregates. With the 
optical observation used we cannot resolve the particles further and do not know the total mass exactly. We used bulk densities instead. In total, the value in eq. \eqref{senergy} should rather be treated as an estimation than an exact result at the moment.
This does not devalue the qualitative dependence of the sticking velocity with water content though.

\section{Conclusion}

In this work we measured the threshold velocity between sticking and bouncing for collisions of ice grains with a solid wall of the same material. We studied three compositions: pure dry ice, pure water ice and a 1 to 1 mixture. As we used the same setup, the results are immediately comparable. In all three cases the average particle size analyzed was $\sim 90$ $\rm \mu m$ $\pm 20$ $\rm \mu m$. Collision velocities were up to 1 m/s. Our goal was to show explicitly how in comparison the sticking properties of these ices differ in collisions. 

While CO$_2$ particles of the given size only stick at velocities below 0.04 m/s, mixtures of 1:1 mass ratio have a sticking threshold at 0.43 m/s. For pure H$_2$O ice we could not find a steep threshold velocity but only probabilities on the order of 60 to 70 \% for sticking up to 1 m/s, which is likely caused by a fragile aggregate structure.
From the sticking velocity for the particles consisting of the mixture of H$_2$O/CO$_2$ ice we calculate the surface energy to $\gamma=2.77^{+0.9}_{-0.8}$ $\text{J}/\text{m}^2$. Compared to pure CO$_2$ ice this is an order of magnitude higher, though the quantitative derivation of this values should be considered with care.

In our earlier paper we showed that CO$_2$ behaves mostly like silicates \citep{Musiolik2016}. 
Here, we show that adding water 
makes particles much stickier. This supports the idea that there is an inner silicate dominated region and and outer CO$_2$ dominated region in protoplanetary disks where collisional growth might be less efficient than in the region in between the respecitve ice lines where water ice dominates.

\section{Acknowledgements}

This work is supported by the DFG under the grant number WU321/12-1 and TE890/1-1.


\end{document}